\documentclass[aps,twocolumn,showpacs,preprintnumbers,amsmath,prl,amssymb,superscriptaddress, a4paper]{revtex4-1}
\usepackage{url}

\usepackage{natbib}
\usepackage{bm}
\usepackage{mathrsfs}
\pdfoutput=1

\usepackage[export]{adjustbox}
\usepackage{amsfonts}
\usepackage{amsmath}
\usepackage[makeroom]{cancel}
\usepackage[english]{babel}
\usepackage[T1]{fontenc}
\usepackage{times}
\usepackage{mathrsfs}
\usepackage{graphicx}% Include figure files
\usepackage{dcolumn}% Align table columns on decimal point
\usepackage{bm}% bold math
\usepackage{wasysym}
\usepackage[colorlinks,bookmarks=true,citecolor=blue,linkcolor=red,urlcolor=blue]{hyperref}
\usepackage[tight, FIGTOPCAP, hang, raggedright, nooneline]{subfigure}
\usepackage{hyperref}
\usepackage{xcolor}
\usepackage{epsfig}
\usepackage{amssymb}
\usepackage{lipsum}
\usepackage{enumitem}

\newcommand*\diff{\mathop{}\!\mathrm{d}}
\usepackage{braket}
\begin{document}
\title{Interaction effects and charge quantization in single-particle quantum dot emitters}
\author{Glenn Wagner}
\affiliation{Rudolf Peierls Centre for Theoretical Physics, Parks Road, Oxford, OX1 3PU, UK}
	\author{Dung X. Nguyen}
\affiliation{Rudolf Peierls Centre for Theoretical Physics, Parks Road, Oxford, OX1 3PU, UK}
\author{Dmitry L. Kovrizhin}
\affiliation{Rudolf Peierls Centre for Theoretical Physics, Parks Road, Oxford, OX1 3PU, UK}
\affiliation{NRC Kurchatov institute, 1 Kurchatov Square, 123182, Moscow, Russia}
\author{Steven H. Simon}
\affiliation{Rudolf Peierls Centre for Theoretical Physics, Parks Road, Oxford, OX1 3PU, UK}

\begin{abstract}
We discuss a theoretical model of an on-demand single-particle emitter that employs a quantum dot, attached to an integer or fractional quantum Hall edge state. Via an exact mapping of the model onto the spin-boson problem we show that Coulomb interactions between the dot and the chiral quantum Hall edge state, unavoidable in this setting, lead to a destruction of precise charge quantization in the emitted wave-packet. Our findings cast doubts on the viability of this set-up as a single-particle source of quantized charge pulses. We further show how to use a spin-boson master equation approach to explicitly calculate the current pulse shape in this set-up.
\end{abstract}
\maketitle

\textit{Introduction.} The venerable field of quantum optics has brought many remarkable technological advances in e.g.~communication and encryption~\cite{Haroche2012}. More fundamentally it has allowed experimental tests of quantum mechanics with unprecedented precision and control. This success would not have been possible without innovations in reliable on-demand single-photon sources. Recently, there has been an exciting new experimental activity in creating and studying analogous sources, but with electrons and fractional quasiparticles in quantum Hall edge states \cite{Parmentier2011,Grenier2011,Parmentier2012,Bocquillon2012,Moskalets2009,Kashcheyevs2012,Moskalets2013,Moskalets2017,Kapfer2018}. The particles emitted by these devices can be entangled using electronic interferometers~\cite{Heiblum2003}, thereby allowing one to extend the ideas developed in quantum optics to the realm of condensed matter physics. More importantly, the particles' statistics are different from that of photons, and they are more amenable to the studies of interaction effects. Therefore, this experimental setting offers new possibilities in manipulating entangled quasiparticle pairs, and in high-precision experimental studies of correlations in many-body electron systems, see review [\onlinecite{Baeuerle2018}]. 

A theoretical proposal for creating coherent single-electron wave-packets from a non-interacting Fermi sea was suggested early on by Levitov et al.~\cite{Leviton,Ivanov:1997}. These works showed that a Lorentzian voltage pulse applied to a one-dimensional conductor can produce a minimal-noise state  having a single 
excited electron and containing no holes at all. This is a desirable feature in the context of electron quantum optics that offers the possibility of using this protocol to design on-demand coherent single-particle electron sources. This work was extended to the case of one-dimensional chiral edge states which arise in the physics of integer and fractional quantum Hall effect (QHE)~[\onlinecite{Keeling:2006}]. The latter set-up has been extensively studied theoretically by Martin et al.,~who focussed on a model of two fractional quantum Hall edge states connected via a quantum point contact (QPC)~\cite{Martin1,Martin2,Martin3,Martin4}. 

\begin{figure}[b]
\includegraphics[width=0.65\columnwidth]{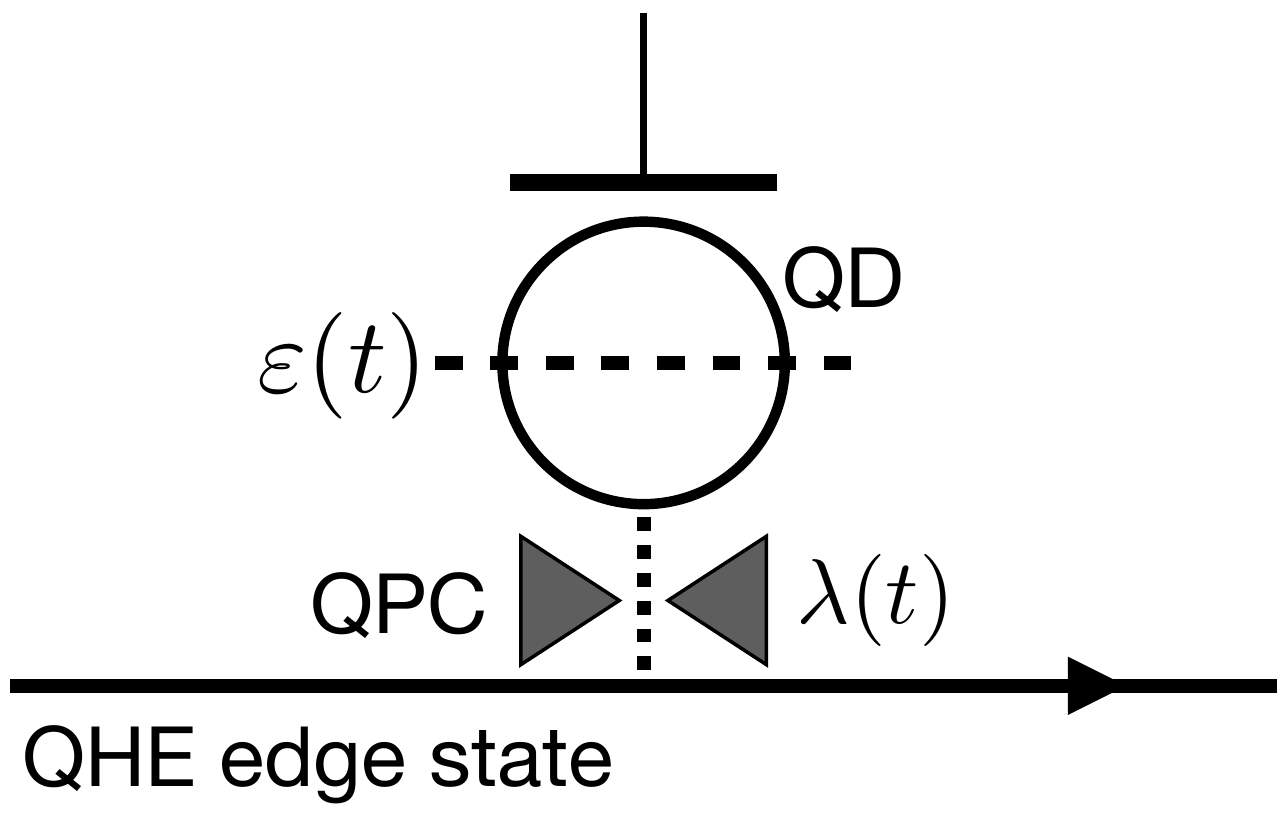}
\caption{Schematic picture of the model. A quantum dot is attached to a quantum Hall edge state (integer or fractional) via a quantum point contact. The voltage on the quantum point contact (QPC) can be used to control the tunneling $\lambda(t)$ between the dot and the edge. We assume that the dot has a single level with energy $\varepsilon(t)$, which is controlled by an applied gate voltage. The level can be occupied with an electron in the IQHE case, or with a quasiparticle in the case of FQHE.}
\label{fig:setup}
\end{figure}

An alternative experimental proposal uses a quantum dot (QD) connected to a quantum Hall edge state~\cite{Glattli2017,Keeling2008,Ferraro2015}. An experiment with such an on-demand single-electron source was performed in \cite{Feve2007}. Here, the putatively quantized pulses are generated via non-equilibrium driving of the quantum dot. In this paper we will study a model of this set-up, shown in Fig.~\ref{fig:setup} with the QD having a single level whose energy can be varied using an applied bias voltage. When this energy rises from below to above the chemical potential a particle can tunnel from the dot into the edge. In the integer quantum Hall effect (IQHE) case a linear voltage-ramp generates a single-electron excitation with minimal noise~\cite{Keeling2008}. The presumed advantage of this set-up is that quantization of charge on the dot is expected to lead to the quantization of the resulting charge pulse on the edge. In contrast, we find that Coulomb interactions, however weak, between the particles on the dot and the edge destroy this precise charge quantization of the emitted current pulse.
  
In this Letter we study the model shown in Fig.~\ref{fig:setup} describing a quantum dot with a time-dependent energy level coupled by tunneling to a chiral QHE edge (integer or fractional). In the integer quantum Hall effect case the energy level on the dot represents an electron, whereas in the fractional (FQHE) case, the energy level may represent either an electron or a fractionally charged quasiparticle. The particle on the dot is allowed to tunnel between the dot and the edge. If the dot contains an electron it may be either inside or outside of the QHE fluid, whereas if it contains a fractionally charged quasiparticle it must be surrounded by FQHE fluid in order to support these fractionalized charges. 

As mentioned above, in both the integer and fractional QHE case, interactions renormalize the charge of the pulses, which can be described within the following physical picture. Due to repulsive Coulomb interactions the charge on the edge close to the QPC gets depleted in the presence of a charge on the dot. Following emission of a particle by the dot, charge fills up the depleted area on the edge, which reduces the net charge flowing downstream from the dot. Therefore, when a particle tunnels from the dot into the edge, while the charge leaving the dot may be quantized, the net charge in the resulting current pulse downstream is always less than the particle charge (for repulsive interactions). 

Our results suggest that creating a source of \textit{precisely} quantized electron or quasiparticle pulses using such a quantum dot set-up would require extra fine-tuning. In order to have a non-vanishing tunneling between the dot and the edge they should be placed in proximity, thus inevitably producing Coulomb interactions between the two. While recent pioneering experiments by Glattli et al.~reported creation of single-electron pulses using a quantum dot set-up in the IQHE case~\cite{Feve2007}, our theory suggests that higher-precision measurements should find that this quantization is not exact, and it would be interesting to compare the results of such measurements with our predictions. In the experimental set-up of \cite{Feve2007}, the Coulomb interactions between the dot and the edge will be partially screened by the metallic gate. However, dipole interactions will remain.

%We would also like to stress that our conclusions are valid in both integer and fractional QHE cases.

The outline of the paper is the following. First we introduce our theoretical model and show how the Hamiltonian of this model can be mapped, via a unitary transformation, to the spin-boson problem. This mapping allows us to analyse the effects of Coulomb interactions between the dot and the edge. In the second part of the paper we use a generalized master equation approach (GME) discussed in \cite{Grifoni1999,Hartmann2000} to obtain results for current pulse profiles. In the Supplementary Material we present a calculation which supports our physical interpretation of the results, as well as a detailed account of the GME approach. We refer to our companion paper for more details on the calculations, where we also compare the results obtained via a spin-boson mapping with the results of perturbative calculations~\cite{Companion}.

\textit{The model.} We consider a theoretical model of the experimental set-up presented in Fig.~\ref{fig:setup}. The model is described by the time-dependent Hamiltonian 
\begin{equation}
\label{eq:Hamiltonian}
	\hat{H}(t)=\hat{H}_0(t)+\hat{H}_{\mathrm{tun}}(t)+\hat{H}_{\mathrm{int}}.
\end{equation}
Here, the first term describes the quantum dot with a single energy level $\varepsilon(t)$ \footnote{In the experimental set-up of Ref. \cite{Feve2007}, the QD has a set of equally spaced energy levels. If the energy level spacing is much larger than the applied bias $\varepsilon(t)$ and the energy scale associated with tunnelling, then we can assume that only the energy level closest to the chemical potential contributes to the dynamics.} which is controlled by a time-dependent gate voltage, together with the edge state with velocity $v$, given in the bosonized form
\begin{equation}
\label{eq:FreeHam}
\hat{H}_0(t)=\varepsilon(t)\hat{S}_z+ \frac{v}{2}\int \frac{dx}{2\pi}\ (\partial_x \hat\varphi)^2.
\end{equation}
Here we introduced spin-1/2 operators describing occupation numbers of the quantum dot, which we treat as a two-level system. The operator $\hat S^+/\hat S^-$ creates/destroys a particle on the QD. In the case of electron tunnelling $\hat S^+$ creates an electron with charge $-e$ on the dot with $e>0$, whereas in the case of a fractionalized charge tunneling it creates a quasiparticle with charge $-\nu e$ \footnote{We note that spin-operators commute with $\hat\psi(x)$ instead of satisfying electron or quasiparticle statistics. However, this does not affect our calculations of the current. See also our companion paper.}. The presence or absence of a particle on the dot is measured by the operator $\hat N= \hat{S}_z+1/2$. In the following we assume large Zeeman splitting and omit the physics of electron spin on the edge. 

The second term in the Hamiltonian (\ref{eq:FreeHam}) describes a chiral edge (for a system of length $L$, assumed very large, with periodic boundary conditions) of a Laughlin state at filling fraction $\nu=1/(2n+1)$, and $n=0,1,2\ldots$ \cite{Wen1992}. 
Here, the bosonic field $\hat{\varphi}$ is given in terms of its eigenmode expansion with momentum $k=2\pi m/L$, $m\in\mathbb{Z}$ as follows \cite{vonDelft1998},
\begin{equation}
\hat{\varphi}(x)=-\sum_{k>0}\sqrt{\frac{2\pi}{k L}}(\hat b_k e^{i k x}
+\hat{b}^{\dagger}_k e^{-ikx}) e^{-ka/2},
\label{eq:eigenmode_expansion}
\end{equation}
where $a$ is the short-distance cutoff, and bosonic operators $\hat{b}_k$ obey commutation relations $[\hat{b}_k,\hat{b}^\dagger_{k'}]=\delta_{kk'}$. Here we will omit zero modes as well as the corresponding Klein factors, as these do not affect the results in the thermodynamic limit in our set-up. We also note that the results do not depend on the cutoff $a$, after sending it to zero at the end of the calculations.  

The electron and quasiparticle operators in the bosonized form \cite{Wen1992,vonDelft1998,vanElburg1998} are described by the vertex operators
\begin{equation}
\label{eq:fermion1}
\hat{\psi}(x)=\frac{1}{\sqrt{2\pi}}a^{-\frac{\gamma^2}{2}}e^{-i\gamma\hat\varphi(x)},
\end{equation} 
where $\gamma=1/\sqrt{\nu}$ for electrons, and $\gamma=\sqrt{\nu}$ for quasiparticles. It is convenient to account for these two different possibilities in a unified manner, and in the following by referring to particles we assume electrons or quasiparticles with the corresponding value of $\gamma$. Note that the charge of the particle is given by $q=-\gamma\sqrt{\nu}e$.

The second term in the Hamiltonian (\ref{eq:Hamiltonian}) describes the coupling of the dot to the edge via a QPC with, in general, time-dependent tunneling amplitude $\lambda(t)$ which can be produced by varying the QPC gate voltage
\begin{equation}
\label{eq:coupling}
\hat{H}_{\mathrm{tun}}(t)=\lambda(t)\hat\psi^\dagger(0) \hat S^-+h.c.
\end{equation}

Finally, we model the Coulomb interactions between the dot and the edge as 
\begin{equation}
\hat{H}_{\mathrm{int}}=-\gamma\frac{g}{2\pi}\partial_x\hat\varphi(0)\hat{S}_z,
\end{equation}
where we used the bosonized form of the charge density operator on the edge $\hat{\rho}(x)=+e\sqrt{\nu}\partial_x\hat\varphi/2\pi$, and $g>0$ being the interaction strength. In this model, the Coulomb interaction is assumed to be a delta-function acting at a single point $x=0$ on the edge. In the case of the Coulomb interaction being spread over a finite region we can still use the above form, where the coupling $g_{\textrm{eff}}$ can be determined from the interaction form, as discussed in the Supplementary Material. 

\textit{Mapping to the spin-boson problem.} One can map \eqref{eq:Hamiltonian} to the well-known spin-boson model using the unitary transformation suggested by Furusaki and Matveev \cite{Furusaki2002} (see also Supplementary Material). Following these authors we define an operator $\hat{U}_1=\exp[-i\gamma\hat\varphi(0)\hat S_z]$. Under a unitary transformation $\hat{\tilde{H}}=\hat{U_1}^\dagger \hat H \hat{U_1}$ the Hamiltonian assumes the spin-boson form which, omitting an unimportant constant, is given by 
\begin{multline}
\hat{\tilde{H}}=\varepsilon(t)\hat S_z + \frac{v}{2}\int\frac{dx}{2\pi}({\partial_x\hat\varphi})^2 \\+\lambda(t)\sqrt{\frac{2}{\pi}} a^{-\frac{\gamma^2}{2}}\hat S_x+v\tilde{\gamma}\hat S_z{\partial_x \hat\varphi(0)}.
\label{eq:Spin_boson_mapping}
\end{multline}
In this representation the effect of the Coulomb interactions amounts to a rescaling of $\gamma$ such that
\begin{equation}
\tilde{\gamma}=\gamma\left(1-\frac{g}{2\pi v}\right).
\label{eq:interactions}
\end{equation}
%We note that the limit of vanishing $\tilde\gamma$ corresponds to complete decoupling of the quantum dot and the edge degrees of freedom in this representation.
After introducing a short-hand notation for the coupling strengths as 
\begin{equation}
\Delta(t)=\lambda(t)\sqrt{\frac{2}{\pi}}a^{-\frac{\gamma^2}{2}},\ \  \eta_k=v\tilde\gamma\sqrt{\frac{2\pi k}{L}}e^{-ka/2}
\label{eq:Delta_Lambda_def}
\end{equation}
we arrive at familiar expression for the spin-boson Hamiltonian c.f.~\cite{Leggett1987},
\begin{multline}
\label{eq:Hamsb}
\hat{\tilde{H}}=\varepsilon(t)\hat S_z + \Delta(t) \hat S_x\\ +\sum_{k>0}\omega_k\hat b^\dagger_k \hat b_k-i\hat S_z\sum_{k>0}\eta_k (\hat  b_k -\hat b^\dagger_k),
\end{multline}
where $\omega_k=v k$. It is worth noting that the transformation between Hamiltonians of Eq.~(\ref{eq:Hamiltonian}) and Eq.~(\ref{eq:Hamsb}) is exact.  

The first two terms of the Hamiltonian in Eq.~(\ref{eq:Hamsb}) represent a spin-1/2 in presence of a time-dependent magnetic field $B(t)=\varepsilon(t)\hat e_z +\Delta(t)\hat e_x$. The last two terms describe the Hamiltonian of a bosonic heat-bath together with the spin-boson coupling. The spectral function of the spin-boson model is defined in the standard way using the following equation
\begin{equation}
\label{eq:spectralJ}
J(\omega)=\pi\sum_{k>0} \eta_k^2 \delta(\omega-\omega_k)=2\pi\alpha\omega\Theta(\omega) e^{-\omega a/v},
\end{equation}
where $\Theta(\omega)$ is the Heaviside theta-function. This corresponds to a heat-bath with Ohmic dissipation, and dimensionless coupling $\alpha=\tilde{\gamma}^2/2$. We estimate for experiments similar to \cite{1a,1b} that $g/2\pi v= 0.04$ and hence $\alpha=0.15$ for the $\nu=1/3$ state. See Supplementary Material for more details. 

\textit{Current.} Now let us turn to a discussion of the main subject of this paper, the behaviour of the current under a non-equilibrium drive of the QD. First, it is useful to obtain general exact results for the current, while we postpone the discussion of the numerical approach to the next section. The Hamiltonian $\eqref{eq:Spin_boson_mapping}$ can be refermionized using a unitary transformation with the operator $\hat{U}_2=\exp[i\tilde\gamma\hat\varphi(0)\hat S_z]$ which brings it into a non-interacting form with the new value of $\tilde\gamma$.  I.e., the Hamiltonian is of the form of Eq.~\eqref{eq:Hamiltonian} except the last term is absent. The equations of motion generated by this Hamiltonian can be used to relate the currents on the edge and on the QD
\begin{equation}
\frac{v}{2\pi}(\partial_x\hat{\varphi}(+0)-\partial_x\hat{\varphi}(-0))=\tilde\gamma\frac{d\hat{N}}{dt}.
\end{equation}
%Since $[\hat U_{1,2},\hat N]=0$, $\hat N$ is unchanged by the transformations allowing us to calculate the current from the transformed Hamiltonian. 
Using equations of motion for $\hat\varphi(x,t)$ away from $x=0$, we obtain an expression for the current on the edge  at $x>0$,
\begin{equation}
\hat{I}(x,t)=-\tilde{q}\frac{d\hat{N}(t-x/v)}{dt},
\label{eq:QP_tunnelling}
\end{equation}
where $\hat{I}=v \hat\rho$ is the current operator, and $\tilde q=(\tilde\gamma/{\gamma})q$. 

One would expect from charge conservation that the proportionality constant should be equal to the charge of the particle $q$. Remarkably, in the interacting case the charge gets renormalized by a factor $\tilde\gamma/{\gamma}$ which is less than one for repulsive interactions. In other words, in the presence of interactions, one cannot obtain a precisely quantized charge pulse.

\textit{Master equation approach.} The mapping to the spin-boson Hamiltonian is particularly useful, since it enables one to use powerful numerical techniques developed for this well-studied problem. For $\alpha<1/2$ one could also use the stochastic Schr\"odinger equation method \cite{Orth2013nonperturbative}. However, in this Letter we will adopt the generalized master equation approach, which makes possible calculations for arbitrary times provided  $\alpha$ is small. This allows the calculation of the current resulting from non-equilibrium driving of the quantum dot. 
 
The starting point of the calculations is the derivation of the path-integral solution for the time-evolution of the reduced density matrix for the spin-1/2 using the Feynman-Vernon influence functional approach, see [\onlinecite{Grifoni1999}]. This is done by exactly tracing out the heat-bath degrees of freedom. From the path-integral solution one then derives the GME describing the time evolution of $\langle \hat{S}_z\rangle$  \cite{Grifoni1999,Hartmann2000},

\begin{equation}
\frac{d}{d t}\langle \hat S_z(t)\rangle=\int_0^t d\tau\ [\frac{1}{2} K^a(t,\tau)-K^s(t,\tau)\langle \hat S_z(\tau)\rangle].
\label{eq:GME}
\end{equation}
Here the integral kernels $K^{(a,s)}(t,\tau)$ can be obtained in terms of a series expansion in $\Delta(t)$ for arbitrary $\alpha$. However, each factor of $\Delta(t)$ in this expansion comes with the integration over time, hence we have to truncate the series in our numerical calculations in the case when $\alpha$ is not small. Remarkably, to linear order in $\alpha$ it is possible to sum up the entire series expansion in $\Delta(t)$ analytically \cite{Hartmann2000} and obtain expressions for $K^{a,s}(t,\tau)$ which are exact in $\Delta(t)$. This truncation of the master equation is useful for $\alpha=\tilde\gamma^2/2\ll1$. %In terms of experiments beyond IQHE set-up, we are especially interested in the fractional case with $\nu=1/3$ being the most robust FQHE state. 
We summarize the derivation of the GME and the definitions of the kernels in the Supplementary Material.

\begin{figure}
    \centering\includegraphics[width=\columnwidth]{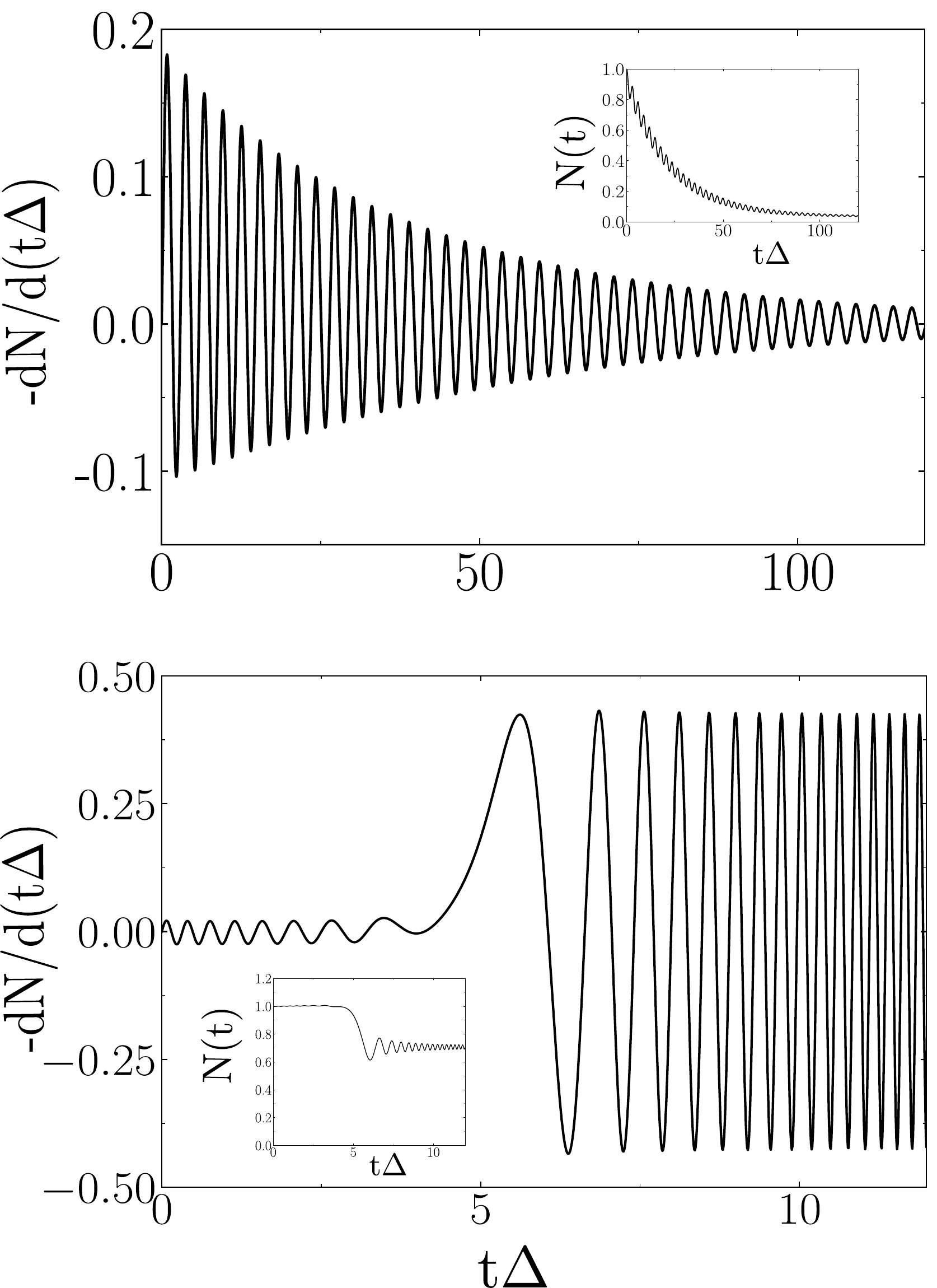}
\caption{Results of the numerical solution of the generalized master equation for the time-dependence of $-d N/d(t\Delta )$, which is related to the current on the edge via Eq.~\eqref{eq:QP_tunnelling}. In both figures we turn-on the tunneling $\lambda(t)$ at the QPC at $t=0$, provided that the dot is filled, and the edge is in equilibrium at $t<0$. (top) Time evolution of the current after a step-like pulse, see text, which leads to discharging of the dot at long times. In the calculations we use parameters $a=0.005v\Delta^{-1}$, $\alpha=0.05$, $\varepsilon_0=2\Delta$. 
(bottom) Time evolution of the current after a linear ramp $\varepsilon(t)=\xi(t-t_0)$ with parameters $a=0.005v\Delta^{-1}$, $\alpha=0.01$, $\xi=4\Delta^2$, $t_0=5\Delta^{-1}$. See insets for the corresponding time-dependence of $N(t)$.}
\label{fig:GME}
\end{figure}

In the top panel of Fig.~\ref{fig:GME} we present the results for the current at constant bias voltage applied to the dot, $\varepsilon(t)=\varepsilon_0$ for $t>0$. The dot is taken to be occupied at $t=0$ corresponding to $\varepsilon(t)$ large and negative for $t<0$. This models the step in the first half-period of a square-wave bias. The time-dependence of the tunneling strength is $\lambda(t)=\lambda\Theta(t)$. Here we use the exact analytical expression obtained in~\cite{Orth2013nonperturbative,Goerlich1989} for the time-evolution, that is valid at $\alpha\ll1$, see details in the Supplementary Material. We find that the current is a highly oscillatory function of time after the voltage ramp and decays exponentially at long times. In the inset we show behaviour of $N(t)$ as a function of time for the same step-function protocol. Notice that the total charge leaving the dot converges to $q$ in the long time limit, which, according to Eq. \eqref{eq:QP_tunnelling}, corresponds to a downstream current pulse of charge $\tilde q$.

In the bottom panel of Fig.~\ref{fig:GME} we present our numerical results using the GME for the current on the quantum dot after a linear voltage ramp with rate $\xi$, so that $\varepsilon(t) = \xi (t-t_0)$. However, in this case in contrast to a step-pulse, not all the charge leaves the quantum dot during the ramp, instead the occupation number of the QD at late times saturates to $\exp(-\pi\Delta^2/2\xi)$. This is rather unexpected because $\varepsilon(t)$ becomes very large at late times. A similar observation was made previously in the context of the spin-boson problem~ \cite{Orth2013nonperturbative,Wubs2006,Saito2007}. This behaviour is distinctly non-adiabatic since the equilibrium occupation of the QD at large bias must vanish. At late times, the current produced by the linear ramp exhibits Rabi oscillations with an instantaneous frequency set by $\varepsilon(t)$~\footnote{Oscillations in the current, when the energy of the dot is well below the chemical potential of the edge (i.e.~before the onset of the voltage ramp), are an artefact of an abrupt switching-on of the tunnelling $\lambda(t)$ at $t=0$.}. 

%At late times $t\Delta \gg1$, not captured in the figure, the current decays exponentially, as in the case of the step-ramp.

In the experimental setting, including effects such as phonons, the remaining charge on the dot is eventually expected to leave the QD at long times, producing a charge pulse downstream with charge $\tilde q$. However, if the current is measured over a timescale shorter than these processes, then our results provide another constraint to quantization of charge pulses in the linear voltage ramp protocol.

\emph{Discussion.} In this paper we studied a theoretical model of a single-particle emitter of charge pulses which uses a quantum dot coupled to a quantum Hall edge state. We showed that it is not possible to obtain precise quantization of these pulses due to Coulomb interactions between the dot and the edge. The interactions effectively add a capacitance to the system, and the charge stored on this capacitor is released in addition to the charge on the dot in the emission process, thus reducing the charge in the outgoing pulse on the edge. Coulomb interactions are unavoidable in the QD set-up, and hence we argue that it is perhaps not the most promising route for creating precisely quantized charge pulses. It would be interesting to compare our theoretical predictions with higher precision measurements of charge in single-particle emitters using a quantum dot, such as in~\cite{Feve2007}.

This raises the question of how to mitigate the destruction of charge quantization if one wants to obtain a single-particle source with precisely quantized charge pulses. In the quantum dot set-up described above, one will want to screen the Coulomb interaction as much as possible in order to minimize the effect, however it can never be eliminated completely.  

Coulomb interactions do not plague proposals where there is no quantum dot but instead a voltage is applied directly to the edge. This makes them perhaps a more promising route to realization of single-particle sources, although the applied voltage pulses must be fine tuned to a Lorentzian profile \cite{Leviton,Jullien2014,Dubois2013}. 

It is also possible to consider a pump geometry \cite{Simon2000,Aleiner,Sharma,Switkes1905}. In this case we must necessarily transfer exactly one quantized charge over one period. However the effect of the interactions is to spread the current over two pulses. There will be a first pulse on the edge as the dot is charged due to the Coulomb repulsion. Then there will be a second pulse when the charge jumps from the dot onto the edge. This second pulse will not carry the full quantized charge due to the depletion of the edge.

From the theoretical perspective we showed how a mapping to the spin-boson problem, and generalised master equation solution can be used to efficiently simulate this interesting class of experimentally-relevant non-equilibrium interacting quantum systems. 

\textit{Acknowledgements.} We would like to thank C. Glattli for insightful discussions and suggestions. We are grateful to J. Keeling, G. Fux, L. Brendon and M. Moskalets for helpful comments on an earlier draft of the paper. This work was supported by EP/N01930X/1. D.K. was supported by EPSRC Grant No. EP/M007928/2. Statement  of  compliance  with  EPSRC  policy  framework  on  research  data:  This  publication is theoretical work that does not require supporting research data.

\bibliography{bib.bib}
	%%%%%%%%%%%%%%%%%%%%%%%%%%%%%%%%%%%%%%%%%%%%%%%%%%%%%%%%%%%%%%%%%%%%%%%%%%%%%

\pagebreak
\clearpage
\begin{widetext}
	\begin{center}
		\textbf{\large --- Supplementary Material ---\\ Interaction effects and charge quantization in single-particle quantum dot emitters}\\
		\medskip
		\text{Glenn Wagner, Dung X. Nguyen, Dmitry L. Kovrizhin and Steven H. Simon }
	\end{center}
    \end{widetext}

	%%%%%%%%%% Merge with supplementary materials %%%%%%%%%%
	%%%%%%%%%% Prefix a "S" to all equations, figures, tables and reset the counter %%%%%%%%%%
	\setcounter{equation}{0}
	\setcounter{figure}{0}
	\setcounter{table}{0}
	\setcounter{page}{1}
	\makeatletter
	\renewcommand{\theequation}{S\arabic{equation}}
	\renewcommand{\thefigure}{S\arabic{figure}}
	\renewcommand{\bibnumfmt}[1]{[S#1]}
	%\renewcommand{\citenumfont}[1]{S#1}
	%%%%%%%%%% Prefix a "S" to all equations, figures, tables and reset the counter %%%%%%%%%%
\section{Estimation of $\alpha$}
The value of alpha can be estimated using results of experiments on electronic Mach-Zehnder interferometers and equilibration of edge states \cite{1a,1b}. Taking from \cite{2a,2b,2c} the values for the effective Fermi-velocity $v=6.5\times 10^4$ m/s, Coulomb energy scale $U=10\mu\textrm{eV}$, and from \cite{3} the length-scale given by the linear size of the quantum dot $l=1\mu$m we obtain an estimate for $g/hv=Ul/hv\sim0.04$ (which is likely to be screened to even smaller values), so that the corrections from interactions which we obtain should be within the error-bars of \cite{Feve2007}. Using this value for the renormalization of gamma, we obtain for the case of quasi-holes in the $\nu=1/3$ case $\alpha=(1/3)\times(1/2)\times(1-0.04)^2=0.15$.

\section{Mean-field theory for finite-size interactions}
\subsection{Effect of time-dependent potential}
In this section, we use a mean-field approach to the Coulomb interactions (where we study instead of a $\delta$-function, a finite size interaction) in which we replace $\hat N(t)$ by $N(t)=\langle \hat N(t)\rangle$. This leads us to investigate the effect of a time-dependent potential on the FQH edge. We study a bosonized Hamiltonian

\begin{equation}
\hat{H}=\frac{v}{4\pi}\int_{-\infty}^\infty(\partial_x\hat\varphi)^2\diff x+\int_{-\infty}^\infty V(x,t)\frac{\partial_x\hat\varphi}{2\pi}\diff x,
\end{equation}
where
\begin{equation}
V(x,t) = \left\{\begin{array}{ll} V(t)=-\frac{\gamma g}{w}(N(t)-\frac{1}{2})  & |x|<w/2  \\  0  &  \textrm{otherwise.} \end{array} \right. 
\label{eq:pot}
\end{equation}
Using the canonical commutation relation $[\hat\varphi(y),\partial_x\hat\varphi]=-2\pi i\delta(x-y)$, the equation of motion for $\varphi=\langle\hat\varphi\rangle$ is
\begin{equation}
\partial_t\varphi(y)+v\partial_y\varphi=-V(y),
\label{eq:EOM}
\end{equation}
which can be solved trivially via Fourier transform to yield
\begin{equation}
\varphi(y,t)=-\int\frac{\diff \omega}{2\pi}\int\frac{\diff k}{2\pi}e^{i(kx-\omega t)}\frac{1}{i}\frac{\tilde V(k,\omega)}{vk-\omega-i\delta},
\label{eq:int}
\end{equation}
where
\begin{equation}
\tilde V(k,\omega)=\int\diff x\int\diff t\ e^{-i(kx-\omega t)}V(x,t).
\end{equation}
Fourier transforming the potential \eqref{eq:pot} and substituting into \eqref{eq:int}
\begin{equation}
\varphi(x,t)=\int\frac{\diff \omega}{2\pi}\frac{\diff k}{2\pi}\tilde V(\omega)e^{-i\omega t}\frac{e^{ikx}(2i\sin kw/2)}{vk-\omega-i\delta}\frac{1}{k},
\label{eq:phi_soln}
\end{equation}
where
\begin{equation}
\tilde V(\omega)=\int\diff t\ e^{i\omega t}V(t).
\end{equation}

\subsection{Solution for $x>w/2$}
Let's consider the $k$ integral first and take $x>w/2$. The semi-circle contour in the upper half plane encloses one pole at $k=\omega/v+i\delta$. There is no pole at $k=0$. 
Substitute the answer into \eqref{eq:phi_soln} to show
\begin{equation}
\varphi(x,t)=\int\frac{\diff \omega}{2\pi}\tilde V(\omega) e^{i\omega (-t+x/v)}\frac{-2\sin \frac{\omega w}{2v}}{\omega+i\delta}
\end{equation}
and hence
\begin{equation}
\partial_x\varphi(x,t)=\frac{1}{v} \bigg[V\bigg(t-\frac{x-\frac{w}{2}}{v}\bigg)-V\bigg(t-\frac{x+\frac{w}{2}}{v}\bigg)\bigg].
\label{eq:discrete_rel}
\end{equation}
For small $w$ we obtain
\begin{equation}
\partial_x\varphi(x,t)=\frac{1}{v^2}w\dot V(t-\frac{x}{v}).
\label{eq:density_potential_relation}
\end{equation}

\subsection{Solution for $|x|<w/2$}
A similar calculation as in the previous section shows that 
\begin{equation}
\partial_x\varphi(x,t)=-\frac{1}{v} V(t-\frac{x}{v})\approx -\frac{1}{v} V(t).
\end{equation}
For a positive charge on the dot, $V(t)>0$ and the Coulomb repulsion means the charge density on the edge is reduced. 

\subsection{Effect of the interactions on the current}
Using translational invariance and \eqref{eq:pot}, \eqref{eq:discrete_rel} becomes
\begin{equation}
\partial_x\varphi(x,t)\bigg\rvert^{x_2}_{x_1}=-\frac{\gamma g}{wv} \bigg(N(t-\frac{x_1}{v})-N(t-\frac{x_2}{v})\bigg).
\label{eq:difference}
\end{equation}
We can add to $\varphi$ any solution that solves the homogeneous equation \eqref{eq:EOM}, ie any right-moving wave solution. Then \eqref{eq:discrete_rel} describes the difference in $\partial_x\varphi$ that we pick up as we pass through a region with potential. 

\begin{figure}[h]
\includegraphics[width=0.5\textwidth]{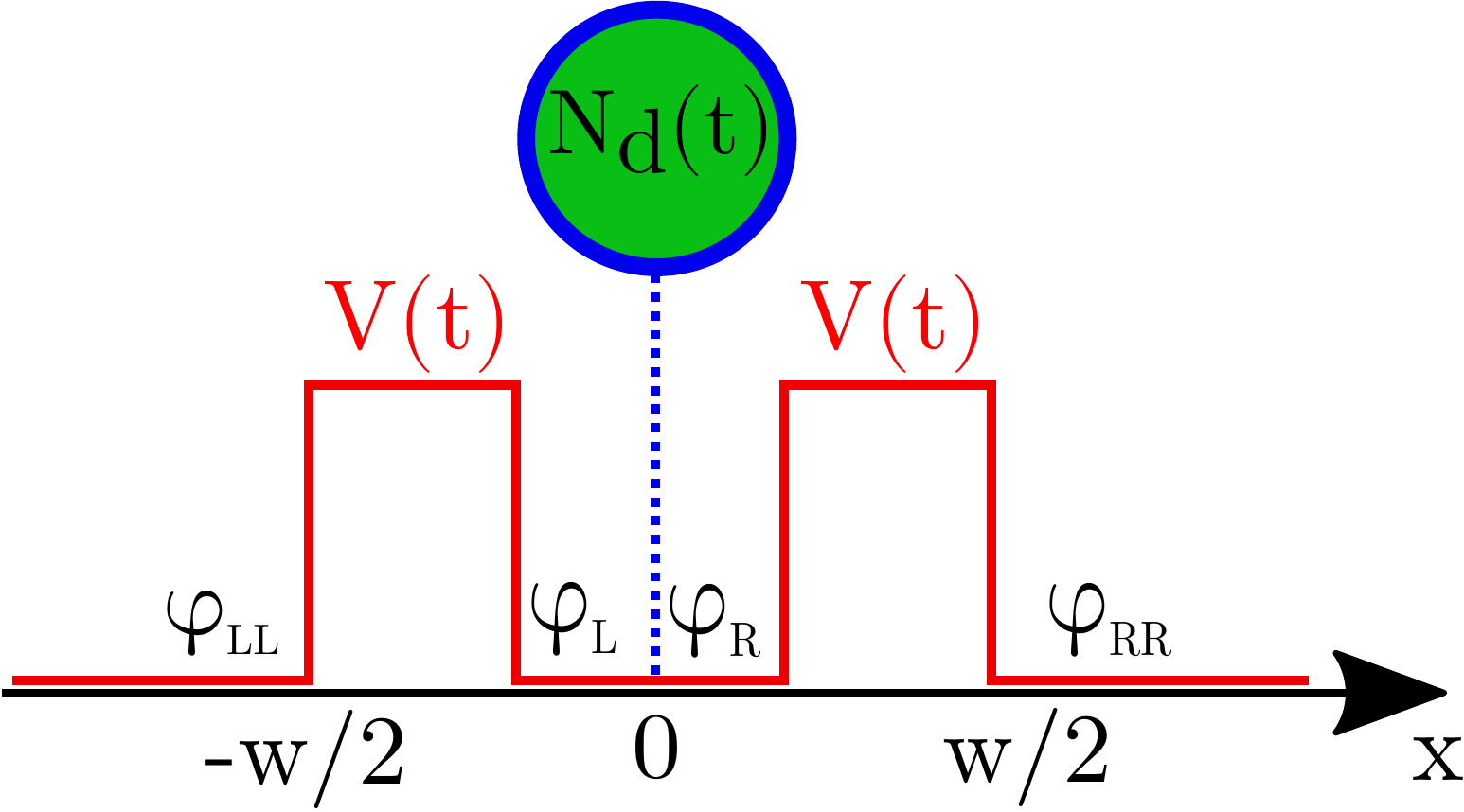}
\caption{Sketch of the potential as a function of position with $\varphi$ defined in four separate regions. Note that the central region where we connect the dot at $x=0$ is assumed to be infinitesimally small.}
\label{fig:potential}
\end{figure}

Now we split our system into various regions as shown in FIG. \ref{fig:potential}. We split the potential into two regions left and right of the contact of width $w/2$ each. Note however, that the final result is independent of the way we split up the region and only depends on the total width of the potential region, $w$. We have normalized the potential such that this final factor of $w$ drops out.

Define $\varphi_{LL}=\varphi(-w/2)$, $\varphi_{L}=\varphi(0^-)$, $\varphi_{R}=\varphi(0^+)$, $\varphi_{RR}=\varphi(w/2)$.
From \eqref{eq:difference} we see that
\begin{equation}
\partial_x\varphi_{RR}-\partial_x\varphi_R=-\frac{\gamma g}{wv} \bigg(N(t)-N(t-\frac{w/2}{v})\bigg).
\label{eq:d_phi1}
\end{equation}
From charge conservation at the contact between the dot and the edge at $x=0$, where we assume there is no potential, we have the current change by an amount
\begin{equation}
\frac{ve\sqrt{\nu}}{2\pi}(\partial_x\varphi_R-\partial_x\varphi_L)=-q\dot N(t),
\label{eq:d_phi2}
\end{equation}
where $q$ is the charge of the particle on the dot. For electrons, $q=-e$ and for quasiparticles $q=-\nu e$. Finally
\begin{equation}
\partial_x\varphi_{L}-\partial_x\varphi_{LL}=-\frac{\gamma g}{wv} \bigg(N(t+\frac{w/2}{v})-N(t)\bigg).
\label{eq:d_phi3}
\end{equation}
Combining \eqref{eq:d_phi1}, \eqref{eq:d_phi2} and \eqref{eq:d_phi3} and expanding for small $w$, the electric current is
\begin{equation}
\frac{ev\sqrt{\nu}}{2\pi}(\partial_x\varphi_{RR}-\partial_x\varphi_{LL})=(-q-e\sqrt{\nu}\frac{\gamma g}{2\pi v}) \dot N(t).
\end{equation}
The first term on the right hand side is simply the usual current we expect to get from the charge on the dot changing with time. The second term on the right is related to charge accumulating on the edge due to the interactions.
\begin{equation}
\frac{ev\sqrt{\nu}}{2\pi}(\partial_x\varphi_{RR}-\partial_x\varphi_{LL})=-q\bigg(1-\frac{g}{2\pi v}\bigg)\dot N(t).
\end{equation}
Using the definition of $\tilde q$ from the main text,
\begin{equation}
\frac{v\sqrt{\nu}}{2\pi}(\partial_x\varphi_{RR}-\partial_x\varphi_{LL}) =-\tilde q\dot N(t).
\end{equation}
This is precisely \eqref{eq:QP_tunnelling} if we set $x=0^+$ and $y=0^-$, which makes sense, since we have taken $w\to 0$. \eqref{eq:QP_tunnelling} was derived from the exact mapping to the spin-boson problem. We thus see that the mean field theory is exact for this calculation. 

\section{Coupling with spatial dependence}
We now consider the most general case, where the coupling between the dot and the edge follows a spatially-dependent profile $g(x)$. In that case
\begin{equation}
V(x,t) = \left\{\begin{array}{ll} -\frac{\gamma g(x)}{w}(N(t)-\frac{1}{2})  & |x|<w/2  \\  0  &  \textrm{otherwise.} \end{array} \right. 
\end{equation}
Choose an infinitesimal $\delta x$ such that we can approximate $g(x)$ as constant in the interval $[x,x+\delta x]$. From \eqref{eq:difference}
\begin{equation}
\label{eq:dphi}
\partial_x\varphi(x,t)\bigg\rvert^{x+\delta x}_{x}=-\frac{\gamma g(x)}{wv^2} \dot N(t)\delta x,
\end{equation}
where we have assumed that $N(t)$ varies on a timescale much longer than $\Omega_c^{-1}=w/v$ so that we can approximate $\dot N(t-x/v)\approx \dot N(t)$. Including higher-order terms in this Taylor expansion leads to higher multipole terms as discussed below. The cut-off frequency is defined as $\Omega_c=v/w$. Summing up all these contributions, we find
\begin{equation}
\partial_x\varphi(x,t)\bigg\rvert^{0^-}_{-w/2}+\partial_x\varphi(x,t)\bigg\rvert^{w/2}_{0^+}=-\frac{\gamma}{wv^2} \dot N(t)\int_{-w/2}^{w/2}g(x)dx.
\end{equation}
This gives us the same result as for the case of the delta-function interaction in the main text or the step-function in the previous section, if we replace $g$ by the effective coupling constant
\begin{equation}
\label{eq:geff}
g_{\textrm{eff}}=\frac{1}{w} \int_{-w/2}^{w/2}g(x)dx.
\end{equation}
There are also multipole contributions to the right hand side of \eqref{eq:dphi}, for example the dipole one with the form 
\begin{equation}
\frac{\gamma}{wv^3} \ddot{N}(t)g(x) x \delta x.
\end{equation}
In comparison with the leading monopole term in \eqref{eq:dphi}, the dipole term is suppressed by order $\omega/\Omega_c$, where $\omega$ is the typical frequency of time dependent $N(t)$. However, this dipole (or higher multipole) term will play an important role when the monopole is screened, in which case $g_{\text{eff}}=0$ in equation \eqref{eq:geff}.  As a consequence, the dipole (and higher multipole) interaction lead to a destruction of a precise charge quantization in the emitted wave-package. 
%%%%%%%%%%%%%%%%%%%%%%%%%%%%%

\section{Details of the master equation}
For the sake of convenience, we summarize results from several previous authors \cite{Grifoni1999,Orth2013nonperturbative,Hartmann2000,Goerlich1989} on numerical solutions to the spin-boson problem in this section. 

We define the Pauli operators by $S_i=\frac{1}{2}\sigma_i$. The Hilbert space of the spin represents a two-level system. Using the Feynman-Vernon influence functional, the authors of [\onlinecite{Grifoni1999}] write down a path-integral solution for the time-evolution of the reduced density matrix of the two-level system  $\rho_{\sigma,\sigma'}(t)$. It is obtained by tracing over the heat bath's degrees of freedom, which can be performed exactly for this system.  One can read off the time evolution of the occupational number $N(t)$ and current profile $I(t)$ from $\rho_{\sigma,\sigma'}(t)$. For a general initial condition $\rho_{\sigma_0,\sigma_0'}(t_0)$ the reduced density matrix evolves as
\begin{equation}
\rho_{\sigma,\sigma'}(t)=\sum_{\sigma_0,\sigma_0'}\int\mathcal{D}\sigma\mathcal{D}\sigma'\mathcal{A}[\sigma]\mathcal{A}^*[\sigma']\mathcal{F}[\sigma,\sigma']\rho_{\sigma_0,\sigma_0'}(t_0)
\end{equation}
where the path integral is over all possible spin paths $\sigma(t)$. $\mathcal{A}[\sigma]$ is the amplitude for the path $\sigma(t)$ when there is no spin-bath coupling. $\mathcal{F}[\sigma,\sigma']$ is the Feynman-Vernon influence functional and captures the effects of the heat bath. This result is exact, however it requires the evaluation of the path integral over all possible spin paths. In practice, the path integral is turned into a sum over spin flips and we integrate over all possible times of the spin flips occurring. In order to evaluate it numerically, this series has to be truncated at a maximum number of spin flips.

The initial condition corresponding to the dot having initial occupation $n_a$ is $\langle\sigma_z\rangle(t=0)=2n_a-1$. Under the initial condition and assuming the spin starts in a pure state, it is shown in \cite{Grifoni1999} that the spin evolves as
\begin{equation}
\langle\sigma_z\rangle=(2n_a-1)P_1^{(s)}(t)+P_1^{(a)}(t),
\label{eq:sigma_z}
\end{equation}
where $P_1^{(s)}(t)$ and $P_1^{(a)}(t)$ are given by a series expansion in $\Delta$. Each factor of $\Delta$ includes an additional time integral, hence limiting the maximum order which we can evaluate numerically. Up to second order in $\Delta$\footnote{Our Hamiltonian differs from that in \cite{Grifoni1999} by a minus sign in the definition of $\varepsilon(t)$ and $\Delta(t)$, hence the expressions below differ in the sign of $\Omega(t)$ from the results in that paper.}
\begin{widetext}

\begin{multline}
P_1^{(s)}(t)=1-\int_0^t\diff t_2\int_0^{t_2}\diff t_1 \Delta(t_2)\Delta(t_1)e^{-Q'(t_2-t_1)}\cos(\Omega(t_1)-\Omega(t_2))\cos\bigg(Q''(t_2-t_1)+Q''(t_1)-Q''(t_2)\bigg),
\end{multline}
and
\begin{multline}
P_1^{(a)}(t)=\int_0^t\diff t_2\int_0^{t_2}\diff t_1 \Delta(t_2)\Delta(t_1)e^{-Q'(t_2-t_1)}\sin(\Omega(t_1)-\Omega(t_2))\sin\bigg(Q''(t_2-t_1)+Q''(t_1)-Q''(t_2)\bigg).
\end{multline}
\end{widetext}
The expansion to second order in $\Delta$ means that we consider paths with at most two spin flips, hence this is an early time approximation valid for $\Delta t\ll1$. For an Ohmic heat bath with spectral function \eqref{eq:spectralJ}, the exact results for $Q'$ and $Q''$ are given in [\onlinecite{Grifoni1997}] and in the limit of large $\omega_c$ and at zero temperature this simplifies to
\begin{equation}
Q'(\tau)=\alpha\ln(1+\omega_c^2\tau^2)
\label{eq:Q'}
\end{equation}
while
\begin{equation}
Q''(\tau)=2\alpha\arctan(\omega_c\tau).
\label{eq:Q''}
\end{equation}
By differentiating \eqref{eq:sigma_z} and using properties of the series expansion of $P_1^{(s)}(t)$ and $P_1^{(a)}(t)$, it can be shown\cite{Grifoni1999} that the spin satisfies the differential equation (GME)

\begin{equation}
\frac{\diff}{\diff t}\langle\sigma_z\rangle_t=\int_0^t\diff t'[K_A^a(t,t')-K_A^s(t,t')\langle\sigma_z\rangle_{t'}]
\label{eq:GME}
\end{equation}
where $K_A^a(t,t')$ and $K_A^s(t,t')$ are also given by a series expansion in $\Delta$. One method to approximate this solution is to expand $K_A^a(t,t')$ and $K_A^s(t,t')$ to lowest order in $\Delta$, this is the so-called non-interacting blip approximation (NIBA), which is valid at early times.

For small $\alpha$ we can sum the entire series expansion in $\Delta$ analytically \cite{Hartmann2000} and obtain expressions for $K_A^a(t,t')$ and $K_A^s(t,t')$ that are exact in $\Delta$ and only require the evaluation of a double integral. From \cite{Hartmann2000}

\begin{multline}
K_A^s(t,t')=\Delta^2\cos[\zeta(t,t')][1-Q'(t-t')]\\
+\int_{t'}^t\diff t_2\int_{t'}^{t_2}\diff t_1\Delta^4\sin[\zeta(t,t_2)]\\
\times P_0(t_2,t_1)\sin[\zeta(t_1,t')][Q'(t-t')\\
+Q'(t_2-t_1)-Q'(t_2-t')-Q'(t-t_1)]
\end{multline}
and
\begin{multline}
K_A^a(t,t')=\Delta^2\sin[\zeta(t,t')]Q''(t-t')\\
-\int_{t'}^t\diff t_2\int_{t'}^{t_2}\diff t_1\Delta^4\sin[\zeta(t,t_2)]
P_0(t_2,t_1)\\
\times \cos[\zeta(t_1,t')][Q''(t-t')-Q''(t_2-t')],
\end{multline}
where $\zeta(t,t')=\int_t^{t'}\varepsilon(s)\diff s$. $P_0(t,t')$ is the solution of \eqref{eq:GME} when there is no spin-bath coupling, ie $\alpha=0$.

This solution is valid to linear order in $\alpha$ but since it is exact in $\Delta$, it is valid at all times. Since it holds well only for $\alpha\ll 1$, it is mainly applicable to the case where we have quasiparticles tunnelling.

To solve the integro-differential equation \eqref{eq:GME} we discretize time into $N$ timesteps and use the method described in \cite{Day1967}. The computation time will scale as $N^4$, thus limiting the latest time up to which we can calculate the current.

We can circumvent this problem for a constant bias. In the case of constant bias $\varepsilon(t)=\varepsilon_0$ and on the same level of approximation as the GME, we have an analytical expression for the current \cite{Orth2013nonperturbative,Goerlich1989}. As derived in \cite{Goerlich1989}, the current decays with a rate 
\begin{equation}
\Gamma=\frac{\pi}{2}\alpha\frac{\Delta_{\textrm{eff}}^2}{\sqrt[]{\Delta_{\textrm{eff}}^2+\varepsilon_0^2}}\ll\Delta,
\end{equation}
where
\begin{equation}
\Delta_{\textrm{eff}}=(\Gamma(1-2\alpha)\cos\pi\alpha)^{1/2(1-\alpha)}\Delta(\Delta/\omega_c)^{\alpha/(1-\alpha)},
\end{equation}
where $\Gamma(x)$ is the gamma function. The current is highly oscillatory with frequency $\Delta_b=\sqrt[]{\Delta_{\textrm{eff}}^2+\varepsilon_0^2}$ and tends to the value $\langle\sigma_z\rangle_\infty=-\varepsilon_0/\Delta_b$ at late times. The time evolution of the spin is 

\begin{multline}
\langle\sigma_z\rangle_t=\langle\sigma_z\rangle_\infty+(\langle\sigma_z\rangle_\infty^2-\langle\sigma_z\rangle_\infty)e^{-2\Gamma t}\\
+\bigg[\frac{\Delta_{\textrm{eff}}^2}{\Delta_{b}^2}\cos\Delta_b t+\bigg(\frac{2\varepsilon_0^2+\Delta_{\textrm{eff}}^2}{\Delta_b^3}-\frac{2\langle\sigma_z\rangle_\infty}{\Delta_b}\bigg)\Gamma\sin\Delta_b t\bigg]e^{-\Gamma t}.
\end{multline}
Numerically, it is found that this approach, called "NIBA+corrections" in \cite{Orth2013nonperturbative} works well for $\alpha<0.05$.

%\end{widetext}

\section{Unitary Transformation}
We start with the Hamiltonian \eqref{eq:Hamiltonian} 
\begin{equation}
	\hat{H}(t)=\hat{H}_0(t)+\hat{H}_{\mathrm{tun}}(t)+\hat{H}_{\mathrm{int}},
\end{equation}
with the components are
\begin{equation}
\hat{H}_0(t)=\varepsilon(t) \hat S^z+ \frac{v}{2}\int_{-L/2}^{L/2} \frac{dx}{2\pi}\ (\partial_x \hat\varphi)^2,
\end{equation}
\begin{equation}
\hat{H}_{\mathrm{tun}}(t)=\lambda(t)\hat\psi^\dagger_{\gamma}(0)\hat{S}^-+h.c.,
\end{equation}
\begin{equation}
\hat{H}_{\mathrm{int}}=-\gamma\frac{g}{2\pi}\partial_x\hat\varphi(0) \hat{S}_z.
\end{equation}
and the fermion operator is defined as \eqref{eq:fermion1}
\begin{equation}
\hat{\psi}_\gamma(x)=\frac{1}{\sqrt{2\pi}}a^{-\frac{\gamma^2}{2}}e^{-i\gamma\hat\varphi(x)}.
\end{equation}
We define the unitary transformation
\begin{equation}
\quad U=e^{i(\tilde\gamma-\gamma)\hat\varphi(0)\hat S_z}, \tilde{\gamma}=\gamma\left(1-\frac{g}{2\pi v}\right).
\end{equation}
Using explicit form of the commutation relations $[\hat\varphi(x),\hat\varphi(x')]=i\pi\textrm{sgn}(x-x')$. One can derive the result of the transformation, omitting an unimportant constant 
\begin{align}
\hat{H}_R=&U^\dagger \hat{H} U\nonumber\\
=&\varepsilon(t)\hat S^z+ \frac{v}{2}\int_{-L/2}^{L/2} \frac{dx}{2\pi}\ (\partial_x \hat\varphi)^2\nonumber\\
&+\lambda(t)a^{\frac{\tilde{\gamma}^2-\gamma^2}{2}}\psi_{\tilde{\gamma}}^\dagger(0)\hat{S}^-+h.c , 
\end{align}
where we used that eigenvalues of $\hat{S}_z$ operator are equal to $\pm1/2$. The result of the unitary transformation removes the interactions. However the parameter $\gamma$ and the tunnelling strength get renormalized.

\section{Plots for further values of $\alpha$}

\begin{figure}
    \centering\includegraphics[width=\columnwidth]{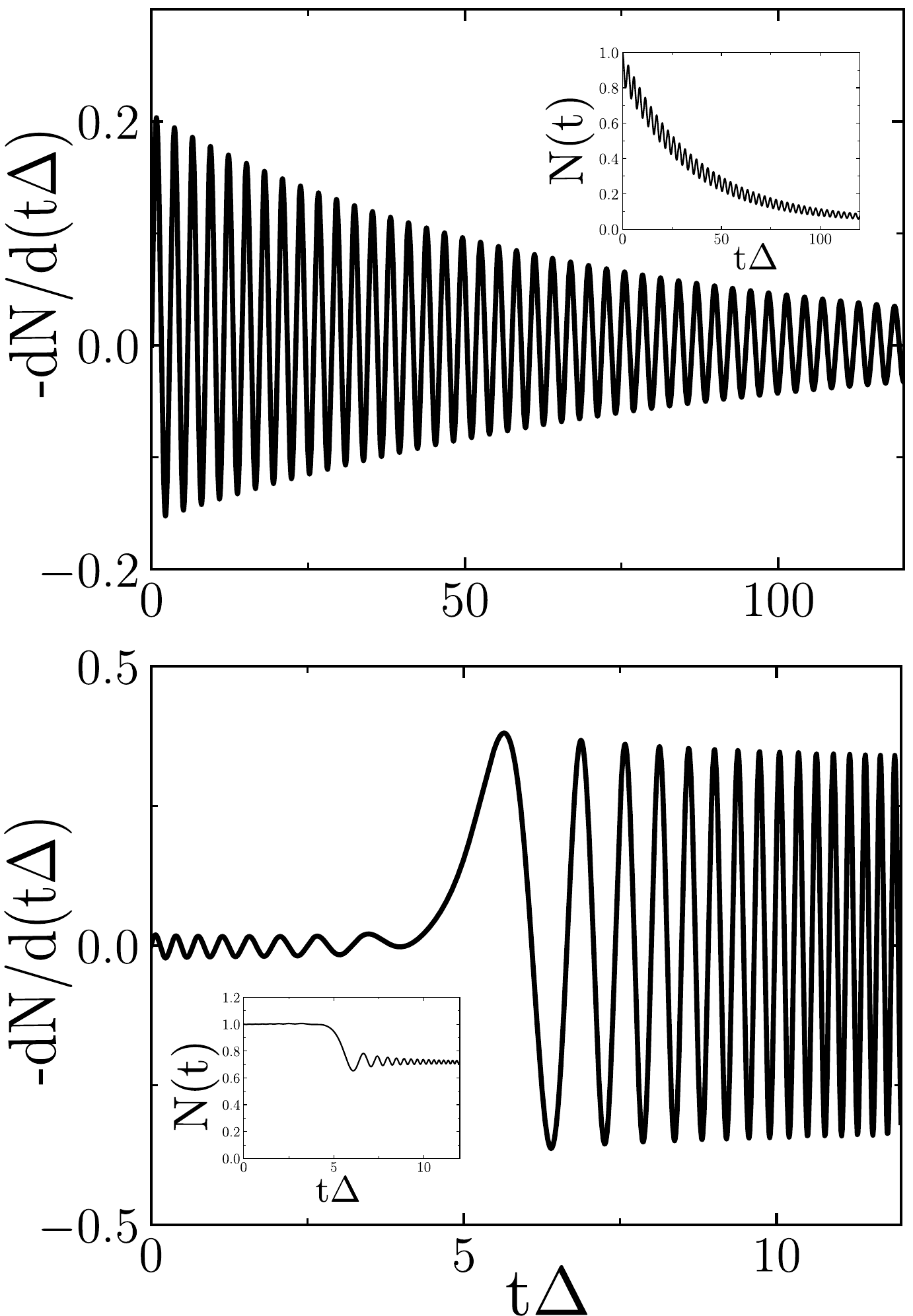}
\caption{Results of the numerical solution of the generalized master equation for the time-dependence of $-d N/d(t\Delta )$, which is related to the current on the edge via Eq.~\eqref{eq:QP_tunnelling}. In both figures we turn-on the tunneling $\lambda(t)$ at the QPC at $t=0$, provided that the dot is filled, and the edge is in equilibrium at $t<0$. (top) Time evolution of the current after a step-like pulse, see text, which leads to discharging of the dot at long times. In the calculations we use parameters $a=0.005v\Delta^{-1}$, $\alpha=0.025$, $\varepsilon_0=2\Delta$. 
(bottom) Time evolution of the current after a linear ramp $\varepsilon(t)=\xi(t-t_0)$ with parameters $a=0.005v\Delta^{-1}$, $\alpha=0.025$, $\xi=4\Delta^2$, $t_0=5\Delta^{-1}$. See insets for the corresponding time-dependence of $N(t)$.}
\label{fig:GME}
\end{figure}

\end{document}